\documentclass[journal]{IEEEtran}
\usepackage{multirow}
\usepackage{amsfonts}
\usepackage{graphicx}
\usepackage{amsmath,amssymb} 
\usepackage{color}
\usepackage[switch]{lineno}

\ifCLASSINFOpdf
\else
\fi
\begin{document}
%
\title{Deep Feature Aggregation and Image Re-ranking with Heat Diffusion for Image Retrieval}
%
%
%

\author{Shanmin~Pang,
        Jin~Ma,
        Jianru~Xue,~\IEEEmembership{Member,~IEEE,}
        Jihua~Zhu,~\IEEEmembership{Member,~IEEE,}
        Vicente~Ordonez
\thanks{S. Pang, J. Ma and J. Zhu are with the School of Software Engineering,
Xi'an Jiaotong University, Xi'an 710049, China.\protect
~E-mail: pangsm@xjtu.edu.cn; m799133891@stu.xjtu.edu.cn;  zhujh@xjtu.edu.cn.}
\thanks{J. Xue is with the Institute of Artificial Intelligence and Robotics (IAIR), Xi'an Jiaotong University, Xi'an 710049, China.\protect
~E-mail: jrxue@xjtu.edu.cn.}
\thanks{V. Ordonez is with the  Dept. of Computer Science, University of Virginia, VA 22904, USA.\protect
~E-mail: vicente@virginia.edu.}}

%
%

\markboth{Journal of \LaTeX\ Class Files,~Vol.~**, No.~**, August~2018}%
{Shell \MakeLowercase{\textit{et al.}}: Bare Demo of IEEEtran.cls for IEEE Journals}
%



\maketitle

\begin{abstract}
Image retrieval based on deep convolutional features has demonstrated state-of-the-art performance in popular benchmarks. In this paper, we present a unified solution to address deep convolutional feature aggregation and image re-ranking by simulating the dynamics of heat diffusion.
A distinctive problem in image retrieval is that repetitive or \emph{bursty} features tend to dominate final image representations, resulting in representations less distinguishable.
We show that by considering each deep feature as a heat source, our unsupervised aggregation method is able to avoid over-representation of \emph{bursty} features. We additionally provide a practical solution for the proposed aggregation method and further show the efficiency of our method in  experimental evaluation.
Inspired by the aforementioned deep feature aggregation method, we also propose a method to re-rank a number of top ranked images for a given query image by considering the query as the heat source.
Finally, we extensively evaluate the proposed approach with  pre-trained and fine-tuned deep networks
on common public benchmarks and show superior performance compared to previous work.
\end{abstract}


%
\IEEEpeerreviewmaketitle

\section{Introduction}
%
%
%
%
\IEEEPARstart{I}mage retrieval has always been an attractive research topic in the field of computer vision.
By allowing users to search similar images from a large database of digital images,
it provides a natural and flexible interface for image archiving and browsing. Convolutional Neural  Networks (CNNs) have shown remarkable accuracy in tasks such as image classification, and object detection. Recent research has also shown positive results of using CNNs
on image retrieval~\cite{babenko2015aggregating,kalantidis2016cross,hoang2017selective, chadha2017voronoi, zhang2018query}. However, unlike image classification approaches which often use global feature vectors produced by fully connected layers, 
these methods extract local features depicting image patches from the outputs of convolutional layers and aggregate these features into \emph{compact} (a few hundred dimensions) image-level descriptors.
Once meaningful and representative image-level descriptors are defined, visually similar images are retrieved by computing similarities between pre-computed database feature representations and query representations.

Thus, a key step contained in these image retrieval methods is to compute global representations.
In order to generate distinguishable image-level descriptors, one has to avoid over-representing bursty (or repetitive) features during the aggregation process.
Inspired by an observation of similar phenomena in textual data, Jegou \textit{et al.} \cite{jegou2009burstiness} identified \emph{intra-image burstiness} as the phenomenon that numerous feature descriptors are almost identical within the same image\footnote{In the original publication \cite{jegou2009burstiness}, the authors define two kinds of burstiness: intra- and inter-image burstiness. Here,  we only study intra-image burstiness, and in what follows it is simplified as burstiness for short.}.
This phenomenon, which appears due to repetitive patterns (e.g., window patches in an image of a building's facade), is widely existed in images containing man-made objects.
Since bursty descriptors are often in large numbers, they may contribute much to the final image representations through inappropriate aggregation strategies such as sum pooling.
However, this is undesirable as such bursty descriptors may correspond to cluttered regions and consequently result in less distinguishable representations.
To address visual burstiness, Jegou \textit{et al.} \cite{jegou2009burstiness} proposed several re-weighting strategies that penalised
descriptors assigned to the same visual word within an image and penalised descriptors matched to many images.
Despite effectiveness, these strategies are only customized for the Bag-of-words (BoW) representation model that considers descriptors individually, and cannot be
combined with \emph{compact} representation models that aggregate local features into a global vector (\emph{i.e.}, do not consider descriptors individually).

In this manuscript, we develop a method to reduce the contribution of bursty features to final image representations in the aggregation stage.
We do so by revealing the relationship among features in an image.
Based on the property of burstiness, the sum of the similarity score between a bursty feature and the whole feature set would be large as there exist many other features identical (or nearly identical) to the bursty feature.
Whereas, the sum of the similarity score between a distinctive feature and the whole feature set tends to be small as it should be  dissimilar to other features.
To formulate this idea, we emulate deep features of an image as a heat system, where the sum of the similarity score is measured as system temperature. Specifically, for a certain feature, we consider it as the unique heat source, and compute the temperature of any other feature with the partial differential equation induced by the heat equation.
Consequently, we define system temperature obtained with this certain feature by summing temperatures of all features.
It is simple to understand that features leading to high system temperatures tend to be bursty ones, while features resulting in low system temperatures are distinctive ones.
Thus, in order to balance the contributions of bursty features and distinctive ones to the final image-level descriptor,
we compel the system temperatures derived from all features (heat sources) in one image to be a constant by introducing a set of weighting coefficients.

Heat diffusion, and more specifically anisotropic diffusion, has been used successfully in various image processing and computer vision tasks. Ranging from the classical work of Perona and Malik~\cite{perona1990scale} to further applications in image co-segmentation, image denoising, and keypoint detection~\cite{zhang2010diffusion,kim2011distributed,karpushin2016keypoint, cho2017geodesic}.
Here, we employ the system temperature induced by this well-known theory for measuring the similarity score between a certain deep convolutional feature and others due to the following two reasons.
First, as well-known, diffusing the similarity information in a weighted graph can measure similarities between different deep features more accurately compared to the pairwise cosine distances.
Second and more importantly, it inspires our second contribution and allows us to obtain considerable performance gains in the image re-ranking stage. Specifically, by considering global similarities that also take into  account the relation among the database representations, we propose a method to re-rank a number of top ranked images for a given query image using the query as the heat source.

Our contributions can be summarized as follows:

\begin{itemize}
  \item Feature weighting: By greedily considering each deep feature as a heat source and enforcing the temperature of the system be a constant within each heat source, we propose a novel efficient feature weighting approach to reduce the undesirable influence of \emph{bursty} features (Section \ref{sec:feature_weighting}).
  \item Image re-ranking: By considering each query image as a heat source, we also produce an image re-ranking technique that leads to significant gains in performance (Section \ref{sec:image_re-ranking}).
  \item We conduct extensive quantitative evaluations on commonly used image retrieval benchmarks, and demonstrate substantial performance improvement over existing unsupervised methods for feature aggregation (Section~\ref{sec:experiments}).
\end{itemize}

The remainder of this manuscript is organized as follows: We briefly overview representative works that close to us in Section~\ref{sec:related_work}.
The details of the proposed feature weighting strategy are described in Section~\ref{sec:feature_weighting}, while the proposed image re-ranking method is presented in Section~\ref{sec:image_re-ranking}. Experimental results are described and discussed in Section~\ref{sec:experiments}, and conclusions are drawn in Section~\ref{sec:conclusions}.

\section{Related Work}\label{sec:related_work}
Since both our deep feature aggregation and image-ranking methods are built on heat diffusion, we therefore first review classical diffusion methods in the computer vision field.
Second, we review representative deep learning based image retrieval methods as this paper also aims to address the image retrieval problem with convolutional neural networks.

\subsection{Diffusion for Computer Vision}
Anisotropic diffusion has been applied to  many computer vision
problems, such as image segmentation~\cite{zhang2010diffusion,kim2011distributed}, saliency detection~\cite{lu2014learning, chen2016discriminative}, and clustering~\cite{donoser2013replicator, pang2018large}.
In these applications, diffusion is used for finding central points by capturing the intrinsic manifold structure of the data.
Our deep feature aggregation method models the problem in the opposite direction, and we differentiate them by weakening deep features that are densely connected to other features with high similarities.

Diffusion is also popular in the context of retrieval~\cite{egozi2010improving, yang2009locally, donoser2013diffusion, furuya2015diffusion, iscen2017efficient}. Among them, the approaches~\cite{egozi2010improving, yang2009locally, furuya2015diffusion} addressed the shape retrieval problem, and performed diffusion on image level. While we focus on the instance-level retrieval problem, and we carry out heat diffusion on deep convolutional features.
Donoser and Bischof~\cite{donoser2013diffusion} reviewed a number of diffusion
mechanisms for retrieval. They focused on iterative solutions
arguing that closed form solutions, when existing, were impractical
due to inversion of large matrices.
However, we rather focus on a closed form solution without iteration as in our case the number of features and the number of re-ranking images is small.
Recently, Iscen \textit{et al.}~\cite{iscen2017efficient} introduced a regional diffusion mechanism on image
regions for better measuring similarities between images. Compared with this method, our re-ranking method is much efficient as we re-rank images based on global level image vectors.

Democratic Diffusion Aggregation (DDA)~\cite{gao2016democratic} is probably the most closest to our feature aggregation method  as it also handled the bursts problem by diffusion.
However, there exists at least a distinctive difference between our method and DDA.
Specifically, we start from the heat equation, and balance the
influence between rare features and frequent ones by enforcing
the system temperatures obtained with different heat sources be a constant.
While DDA inherited from Generalized Max-Pooling (GMP)~\cite{murray2017interferences}, which equalized the contribution of a single descriptor to the aggregated vector.
Furthermore, we provide a unified solution to feature aggregation and image re-ranking, which is otherwise not possible by \cite{gao2016democratic}.

\subsection{Deep Learning for Image Retreival}
Early attempts to use deep learning for image retrieval considered the use of the activations of fully connected layers as image-level descriptors~\cite{babenko2014neural,razavian2014cnn,gong2014multi}.
In~\cite{babenko2014neural}, a global representation was derived from the output of the penultimate layer.
This work was among the first to show better performance than traditional methods based on SIFT-like features at the same dimensionality.
Concurrently, Gong \textit{et al.}~\cite{gong2014multi} extracted multiple fully-connected activations by partitioning images into fragments, and then used VLAD-embeddings~\cite{jegou2010aggregating} to aggregate the activations into a single image vector.
The work~\cite{razavian2014cnn} reported promising results using sets of a few dozen
features from the fully-connected layers of a CNN, without aggregating them into a global descriptor. However, observing that neural activations of the lower layers of a CNN capture more spatial details, later works advocated using the outputs of convolutional layers as features~\cite{babenko2015aggregating,kalantidis2016cross,hoang2017selective,tolias2016particular,azizpour2015generic,razavian2016visual,noh2017largescale,xu2018unsupervised, pang2018building}.
These convolutional features were subsequently used for similarity computation either with individual feature matching~\cite{noh2017largescale} or with further aggregation steps~\cite{babenko2015aggregating,kalantidis2016cross,hoang2017selective,tolias2016particular}.
In this work, we consider convolutional features as local features, and
aggregate them into a global image descriptor.

Considerable effort has been dedicated to aggregating the activations of convolutional layers into a distinctive global image vector.
For instance, \cite{azizpour2015generic,razavian2016visual} evaluated image-level descriptors obtained using max-pooling over the last convolutional layer, while Babenko and Lempitsky~\cite{babenko2015aggregating} showed that sum-pooling leads to better performance.
Kalantidis \textit{et al.}~\cite{kalantidis2016cross} further proposed a non-parametric method to learn weights for both spatial locations and feature channels.
Related to that, Hoang \textit{et al.}~\cite{hoang2017selective} proposed several masking schemes to select a representative subset of local features before aggregation,
and achieved satisfactory results by taking advantage of the triangulation embedding~\cite{jegou2014triangulation}.
Similarly, we proposed a deep feature selection and weighting method using the replicator equation in our very recent work~\cite{pang2018building}.
In another work, Tolias \textit{et al.}~\cite{tolias2016particular} computed a collection of region vectors with max-pooling on the final convolutional layer, and then combined all region vectors into a final global representation.
More recently, Xu \textit{et al.}~\cite{xu2018unsupervised} independently employed selected part detectors to generate regional representations with weighted sum-pooling~\cite{kalantidis2016cross}, and then concatenated regional vectors as the global descriptor. In this paper, we instead propose heat diffusion to weight and then aggregate deep feature descriptors.

Fine-tuning an off-the-shelf network is also popular for improving retrieval quality.
For instance, there are a number of approaches that learn features for the specific task of landmark retrieval~\cite{radenovic2016cnn,gordo2016deep,arandjelovic2016netvlad,cao2016quartet}.
While fine-tuning a pre-trained model is usually preceded by extensive manual annotation,
Radenovic \textit{et al.}~\cite{radenovic2016cnn} introduced an unsupervised fine-tuning of CNN for image retrieval from a large collection of unordered images in a fully automated manner.
Similar to this work, the methods presented in~\cite{arandjelovic2016netvlad,cao2016quartet} overcame laborious annotation, and collected training data in a weakly supervised manner.
More specifically, Arandjelovic \textit{et al.}~\cite{arandjelovic2016netvlad} proposed a new network architecture, NetVLAD, that was trained for place recognition in an end-to-end manner from weakly supervised Google Street View Time Machine images. Cao \textit{et al.}~\cite{cao2016quartet} trained a special architecture Quartet-net by harvesting data automatically from GeoPair~\cite{thomee2016yfcc100m}. We show that our feature weighting, and image re-ranking approach, while not requiring extra supervision, performs favorably compared to these previous methods.

In a couple of very recent works~\cite{iscen2017efficient,iscen2017fast},
images were represented by multiple high-dimensional regional vectors.
These two approaches achieve great performance on common benchmarks, they are however computationally demanding, both in terms of
memory and computational usage. In contrast, our work uses a single vector representation while achieving similar performance.

\section{Feature Weighting with the Heat Equation}\label{sec:feature_weighting}
Given an input image $I$ that is fed through a pre-trained or a fine-tuned CNN,
the activations (responses) of a convolutional layer form a 3D tensor
$\boldsymbol{X} \in \mathbb{R}^{W\times H \times K}$, where $W \times H$ is the spatial resolution of the feature maps,
and $K$ is the number of feature maps (channels).
We denote $\mathcal {V}= \left\{\boldsymbol{f}_{l}\right\}$ as a set of $W \times H$ local features,
where $\boldsymbol{f}_{l}$ is a $K$-dimensional vector  at spatial location $(i,j)$  in $\boldsymbol{X}$.
That is to say,~$l = i+(j-1)\times W$, where $1 \leq i \leq W$ and $1 \leq j \leq H$.
We assume that Rectified Linear Units (ReLU) are applied as a last step,
guaranteeing that all elements of $\boldsymbol{f}_{l}$ are non-negative.

\subsection{Problem Formulation}
We utilize the theory of anisotropic diffusion~\cite{weickert1998anisotropic}
to compute weights for each feature in $\mathcal {V}$  based on their distinctiveness, thus avoiding
the burstiness issue. Let us assume that the deep feature point set $\mathcal {V} = \{\boldsymbol{f}_l\}$ constitute an undirected graph. By assigning $\boldsymbol{f}_l$ as the unique heat source, we assume that the graph constitutes a  heat transfer system, and the \emph{linear heat equation} for this system is defined as follows~\cite{weickert1998anisotropic}:
\begin{equation}
\partial_t\boldsymbol{\mu}_l = \text{div}(\boldsymbol{P}\nabla \boldsymbol{\mu}_l),
\end{equation}
where $\boldsymbol{\mu}_l = (\mu_l(1,t), \mu_l(2,t),\ldots, \mu_l(|\mathcal {V}|,t))^{T}$, and its $m$-th entry $\mu_l(m,t)$ represents the temperature at node $\boldsymbol{f}_m$ at time $t$.
$\boldsymbol{P} = (P(m,n))_{|\mathcal {V}|\times |\mathcal {V}|}$ is a positive definite symmetric matrix
called the \emph{diffusion tensor},  and $P(m,n)$ denotes the $(m,n)$-th diffusion coefficient reflecting the interactions between the feature pair $\boldsymbol{f}_m$ and $\boldsymbol{f}_n$.

Our problem is to compute the temperature at each node $\boldsymbol{f}_m$.
That is,
\begin{equation}\label{heat_equation1}
\left\{
\begin{array}{ll}
\frac{\partial \mu_{l}(m,t)}{\partial t}= \text{div}(\boldsymbol{P}\nabla \mu_l(m,t)),\\
s.t.~\mu_l(g)=0,~\mu_l(l)=1,
\end{array}
\right.
\end{equation}
where we use the \emph{Dirichlet} boundary conditions~\cite{weickert1998anisotropic},
and assume that the temperature of the environment node (outside of the system) and the source node
is always zero (i.e., $\mu_l(g)=0$) and one (i.e., $\mu_l(l)=1$), respectively.

In practice, we compute the temperature at each node with the following simplified assumptions.
Specifically, we let $t\rightarrow+\infty$ and consequently drop $t$ in our method
as we are interested in the steady state, and define the diffusion tensor $\boldsymbol{P}$
by the \emph{cosine} similarity between deep feature vectors:
\begin{equation}
 P(m,n)= \left\{
\begin{array}{ll}
0, & m=n,\\
\frac{\boldsymbol{f}_m^{T}\boldsymbol{f}_n}{\|\boldsymbol{f}_m\|\|\boldsymbol{f}_n\|}, & m\neq n.
\end{array}
\right.
\end{equation}
Furthermore, we assume that the dissipation heat loss at a node $\boldsymbol{f}_m$ is $\lambda_m$, which is constant in time.
In other words, each node $\boldsymbol{f}_m \in \mathcal {V} \setminus f_l$ is connected to an environment node $g$ with diffusivity of $\lambda_m$.
With these assumptions, the heat diffusion Eq.(\ref{heat_equation1}) reduces to the simplified version~\cite{weickert1998anisotropic,grady2006random}:
\begin{equation}\label{heat_equation2}
\left\{
\begin{array}{ll}
\mu_l(m)= \frac{1}{a_m}\sum P(m,n)\mu_l(n),\\
s.t.~\mu_l(g)=0,~\mu_l(l)=1,
\end{array}
\right.
\end{equation}
where $a_m = \sum P(m,n) + \lambda_m,~m \neq g, l$. Without loss of generality, we assume $\boldsymbol{P} = (0, \boldsymbol{P}_1^{T}; \boldsymbol{P}_1, \boldsymbol{P}_2)$,
where $\boldsymbol{P}_1 \in \mathbb{R}^{(|\mathcal {V}|-1)\times 1}$ stores the similarities between points in $\mathcal {V} \setminus \boldsymbol{f}_l$ and $\boldsymbol{f}_l$,  and $\boldsymbol{P}_2 \in \mathbb{R}^{(|\mathcal {V}|-1)\times (|\mathcal {V}|-1)}$
stores the similarities between any two pair points in $\mathcal {V} \setminus \boldsymbol{f}_l$.
Then, Eq.(\ref{heat_equation2}) can be rewritten as
\begin{equation}\label{heat_equation3}
\boldsymbol{\mu}_l = (\boldsymbol{I}_{|\mathcal {V}|-1}-\boldsymbol{\Lambda}^{-1}_{l}\boldsymbol{P}_2)^{-1}(\boldsymbol{\Lambda}^{-1}_{l}\boldsymbol{P}_1),
\end{equation}
where $\boldsymbol{I}_{|\mathcal {V}|-1}$ is the identity matrix of size $(|\mathcal {V}|-1) \times (|\mathcal {V}|-1)$, and $\boldsymbol{\Lambda}_l = \text{diag}(a_1,\ldots, a_{l-1}, a_{l+1}, \ldots, a_{|\mathcal {V}|})$ is the diagonal matrix.
Thus, the temperature of the system induced by the heat source $\boldsymbol{f}_l$ is defined as
\begin{equation}\label{psi_l}
\psi_l = \sum_{m=1}^{|\mathcal {V}|}\mu_l(m).
\end{equation}

\begin{figure}[t]
    \centering
    \includegraphics[width=1.7in]{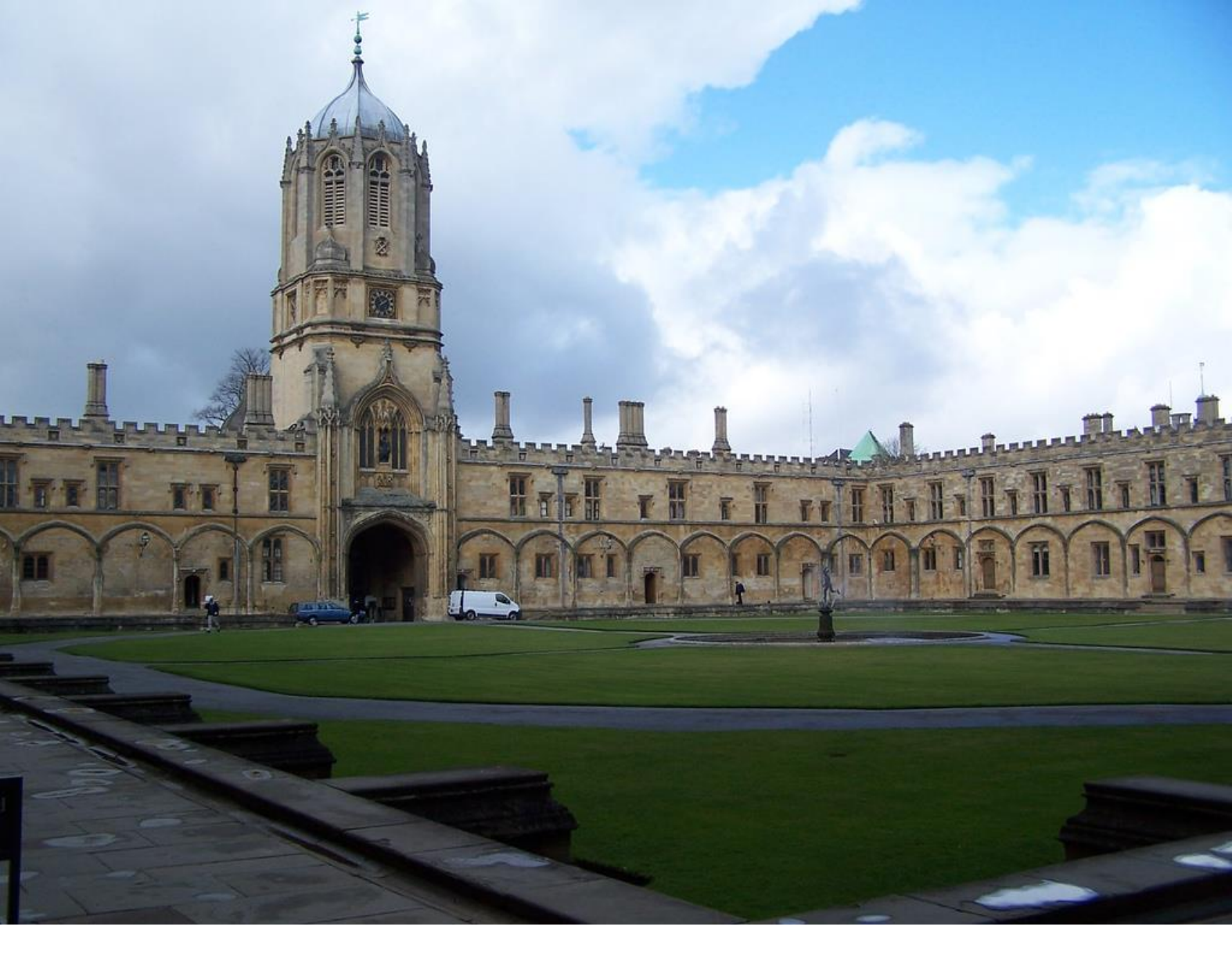}
    \includegraphics[width=1.72in]{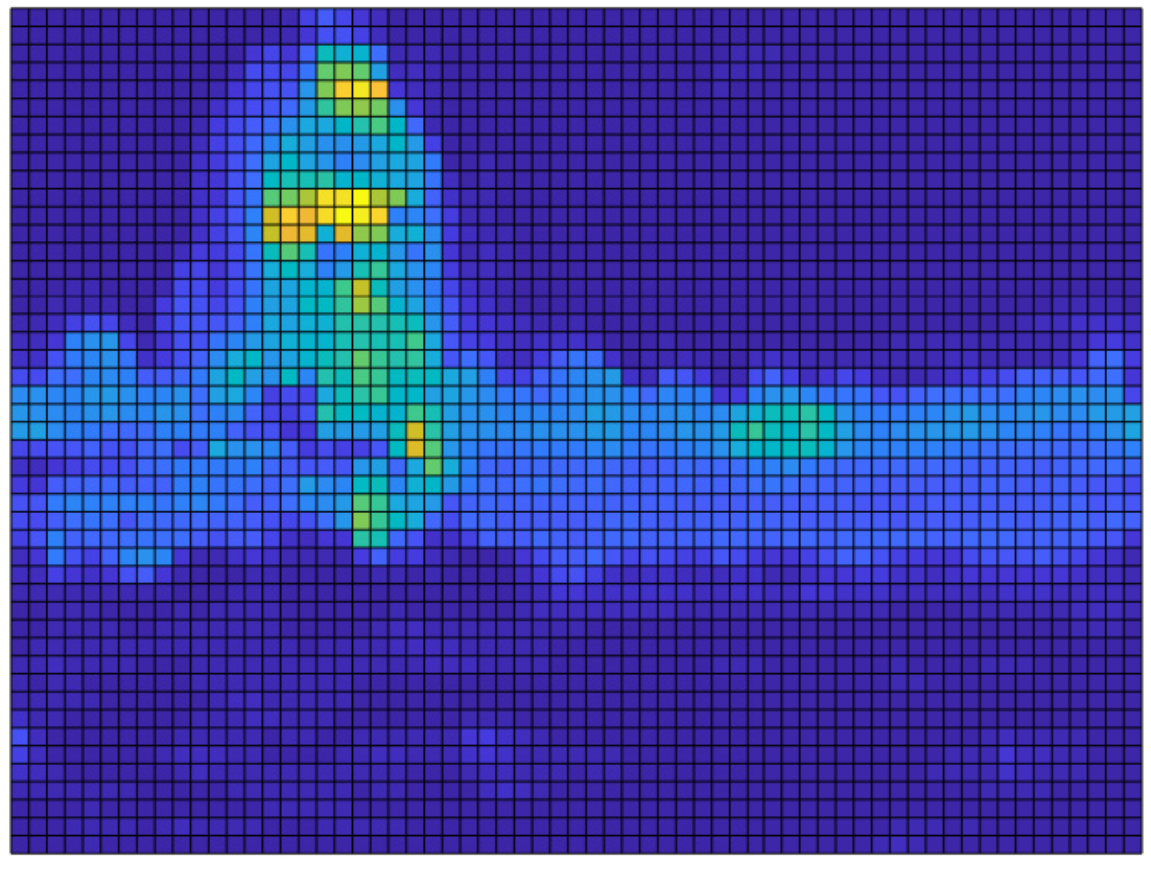}
    \includegraphics[width=1.7in]{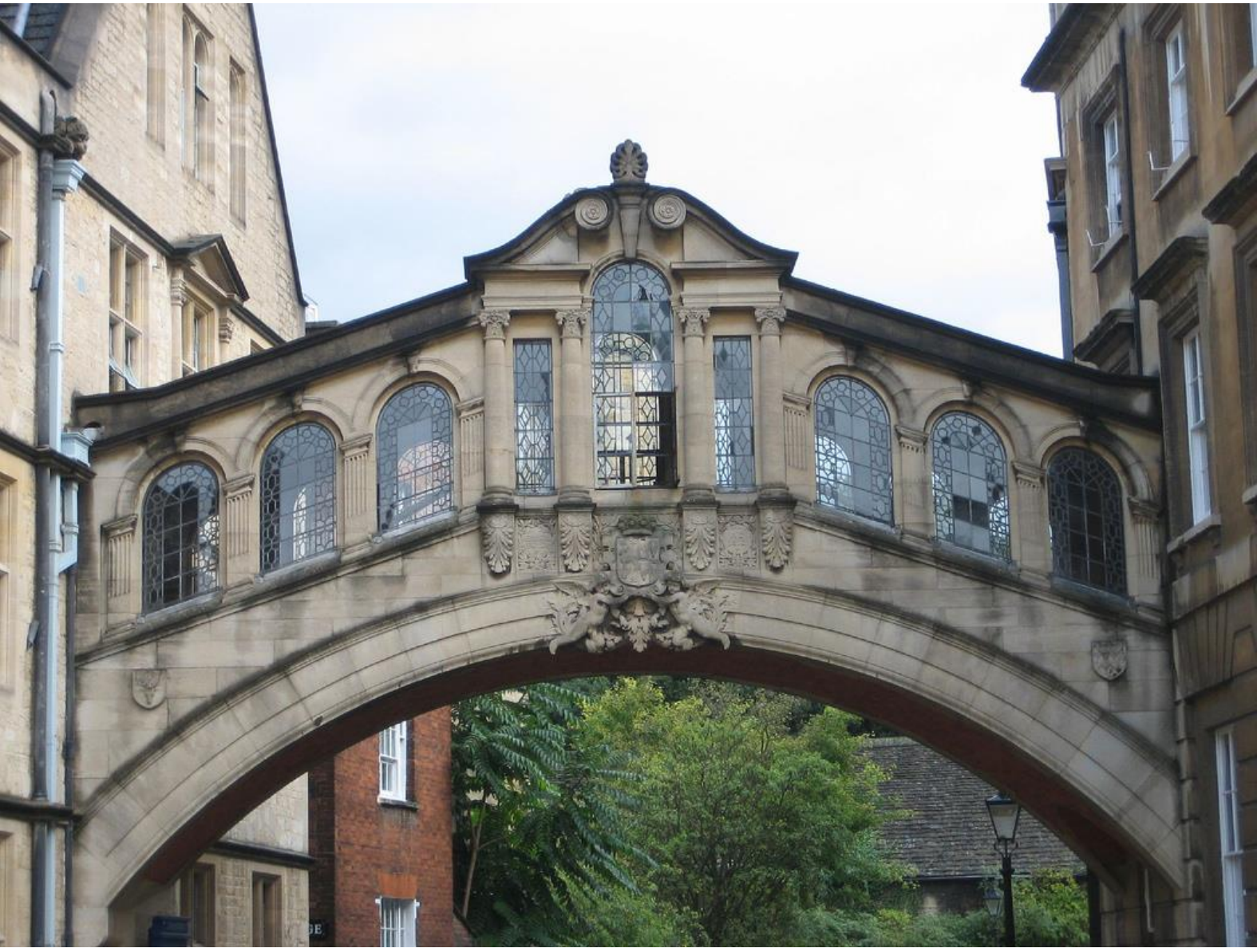}
    \includegraphics[width=1.73in]{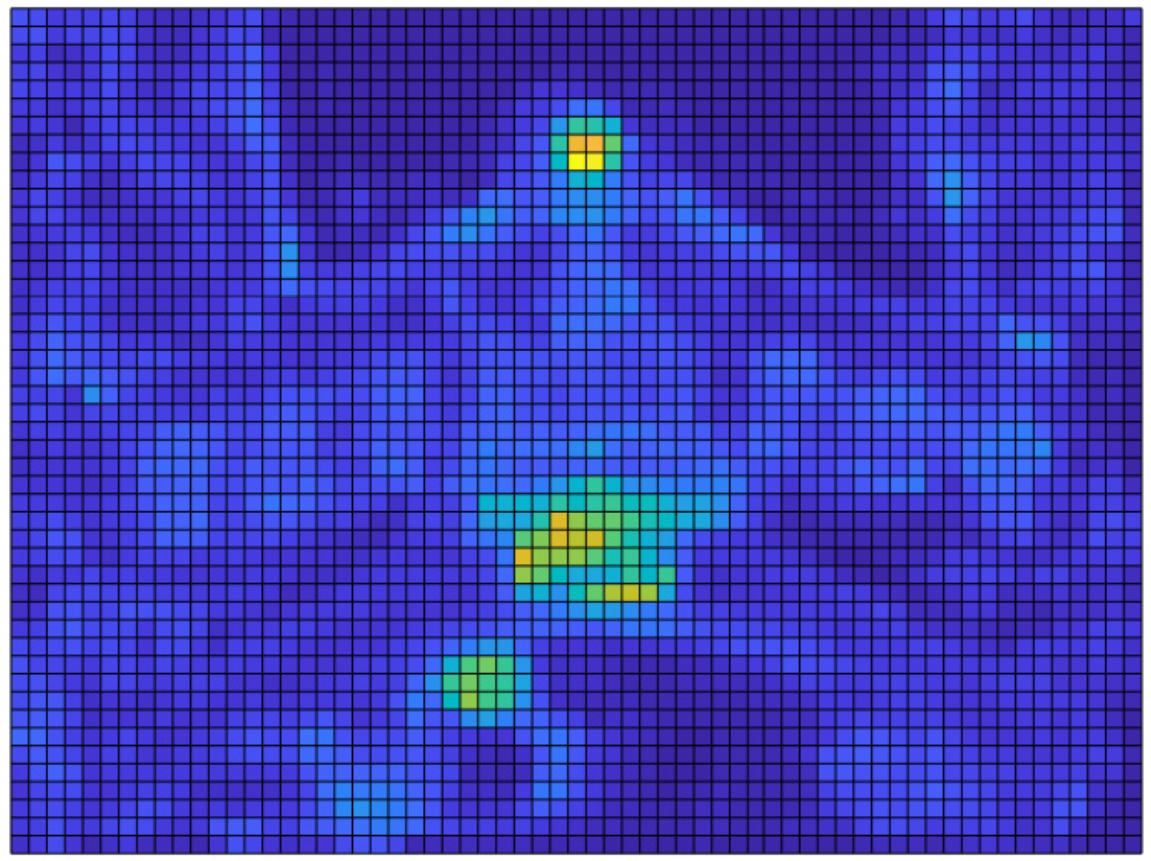}
    \caption{Two example visualization results of the proposed method with deep features extracted using the SiaMAC$\dagger$~\cite{radenovic2016cnn} CNN model. Left: the original image; Right: relative weights (warmer colors indicate larger weights) for deep convolutional features. For instance, in the top image,
    our feature weighting method assigns larger weights to features corresponding to the distinctive area of tower, and smaller weights to ones corresponding to the repetitive areas of grass and sky.}\label{fig:visualizing_results}
\end{figure}

We greedily consider each point $\boldsymbol{f}_l \in \mathcal {V}$ as a heat source, and compute the corresponding system temperature $\psi_l$ under the linear anisotropic diffusion equation.
The value of $\psi_l$ can indicate whether $\boldsymbol{f}_l$ is a bursty feature or a distinctive one.
As described earlier, a bursty feature tends to be identical (or nearly identical) to many other features, whereas a distinctive feature is prone to be dissimilar to other features.
This means a bursty feature is densely connected to other features with high similarities, which consequently rises the temperature of the system.
In contrast, a distinctive feature connects other features sparsely and therefore causes the system temperature to be low.
Thus, in order to balance the influence between bursty features and distinctive ones,
we compel the system temperatures derived from all features (heat sources) in one image to be a constant by introducing a set of weighting coefficients $w_l$.
That is,
\begin{equation}
w_l\times \psi_l = 1, ~\forall \boldsymbol{f}_l \in \mathcal {V}.
\end{equation}

As a result, $w_l$ is used to reduce the burstiness effects,
and we accordingly compute the final image representation $\boldsymbol{v}_I$ of each image $I$ by
\begin{equation}\label{representation_equation}
\boldsymbol{v}_I = \frac{\left(\sum_{l=1}^{|\mathcal {V}|}w_l\boldsymbol{f}_l\right)^{\alpha}}{\left\|\left(\sum_{l=1}^{\mathcal {V}}
w_l\boldsymbol{f}_l\right)^{\alpha}\right\|},
\end{equation}
where $0 < \alpha \leq 1$ is a constant, and we typically set $\alpha = 0.5$.~$\alpha$ plays the same role as the exponent parameter in the Power-law Normalization (PN) formula~\cite{perronnin2010improving}.
However, it is worth noting that, we apply $\alpha$-normalisation on the image vector \emph{before} PCA whitening,
while PN is integrated in the retrieval frameworks~\cite{murray2017interferences,do2017embedding}
\emph{after} rotating the image representation with a PCA rotation matrix. Fig.~\ref{fig:visualizing_results} visualizes the weights computed for two sample input images, larger weights are shown in warmer colors. As shown, our feature weighting method assigns larger weights to distinctive areas, and smaller weights to repetitive ones.

For convenience, in the following we denote our heat equation based feature weighting method presented in this section as HeW.

\subsection{Computing Weights in Practice}
It seems that we have to solve Eq.(\ref{heat_equation3}) $|\mathcal {V}|$ times to get the image representation of $I$, and each time
we need to solve a linear equation of size $(|\mathcal {V}|-1)\times (|\mathcal {V}|-1)$.
Thus, the total time cost of HeW is about in $O(|\mathcal {V}|^4)$.
This might be computationally intensive if the selected feature set cardinality $|\mathcal {V}|$ is large.

However, the actual time complexity can be reduced to $O(|\mathcal {V}|^3)$,
and we can compute all $\psi_l$ by inverting the matrix  $(\boldsymbol{I}_{|\mathcal {V}|}-\boldsymbol{\Lambda^{-1} P})$ only once, where $\boldsymbol{\Lambda} = \text{diag}(a_1,\ldots,  a_{|\mathcal {V}|})$.
Specifically, we take computing $\psi_{1}$ as an example to illustrate the practical computational process.
We leverage the block structure of $(\boldsymbol{I}_{|\mathcal {V}|}-\boldsymbol{\Lambda^{-1} P})$, i.e.,
\begin{equation}
\boldsymbol{I}_{|\mathcal {V}|}-\boldsymbol{\Lambda^{-1} P} =
{\left[ \begin{array}{cc}
1 & -\boldsymbol{x}^{T}\\
-\boldsymbol{y} & \boldsymbol{Q}
\end{array}
\right]},
\end{equation}
where $\boldsymbol{x}$ and $\boldsymbol{y}$ are vectors of size $|\mathcal {V}|-1$, and $\boldsymbol{Q}$ is the matrix of size
$(|\mathcal {V}|-1)\times (|\mathcal {V}|-1)$. According to Eqs. (\ref{heat_equation3}) and (\ref{psi_l}),
it is simple to know that
\begin{equation}
\psi_1 = \sum\boldsymbol{Q}^{-1}\boldsymbol{y} +1.
\end{equation}
By leveraging the property of the block matrix, we can derive that  $(\boldsymbol{I}_{|\mathcal {V}|}-\boldsymbol{\Lambda^{-1} P})^{-1} =$
\begin{equation}
{\left[ \begin{array}{cc}
1+\boldsymbol{x}^{T}(\boldsymbol{Q}-\boldsymbol{yx}^{T})^{-1}\boldsymbol{y}~&~ \boldsymbol{x}^{T}(\boldsymbol{Q}-\boldsymbol{yx}^{T})^{-1}\\
(\boldsymbol{Q}-\boldsymbol{yx}^{T})^{-1}\boldsymbol{y} ~&~ (\boldsymbol{Q}-\boldsymbol{yx}^{T})^{-1}
\end{array}
\right]}.
\end{equation}
Furthermore, one can prove that
\begin{equation}
\boldsymbol{Q}^{-1}\boldsymbol{y} = \frac{(\boldsymbol{Q}-\boldsymbol{yx}^{T})^{-1}\boldsymbol{y}}{1+\boldsymbol{x}^{T}(\boldsymbol{Q}-\boldsymbol{yx}^{T})^{-1}\boldsymbol{y}}.
\end{equation}
The above three equations
demonstrate that we can derive $\psi_1$ by using the first column of $(\boldsymbol{I}_{|\mathcal {V}|}-\boldsymbol{\Lambda^{-1} P})^{-1}$.
Similarly, we can get $\psi_l$ using the $l$-th column of $(\boldsymbol{I}_{|\mathcal {V}|}-\boldsymbol{\Lambda^{-1} P})^{-1}$:
\begin{equation}
\psi_l = \frac{\sum_{m=1,m\neq l}^{|\mathcal {V}|}(\boldsymbol{I}_{|\mathcal {V}|}-\boldsymbol{\Lambda^{-1} P})^{-1}(m,l)}{(\boldsymbol{I}_{|\mathcal {V}|}-\boldsymbol{\Lambda^{-1} P})^{-1}(l,l)}+1.
\end{equation}
Thus, the conclusion that the computational cost is in $O(|\mathcal {V}|^3)$ holds.

For very large-scale image retrieval, the time complexity in  $O(|\mathcal {V}|^3)$ might still be computationally intensive as we need to process too many images. However, it is worth noting that database representations are \emph{pre-computed} in a retrieval system, and therefore image representation efficiency is not so important.
Furthermore, as each database image can be dealt with independently, one can take advantage of the parallel processing technique (with a multi-core computer or with multiple computers, or even with both) for fast computing image-level descriptors of database images.

\section{Image Re-ranking with the Heat Equation}\label{sec:image_re-ranking}
Inspired by our deep feature aggregation method HeW, we propose a heat equation based re-ranking approach HeR in this section.
Given a query image $q$, we denote its image vector produced by HeW or other potential image representation methods as $\boldsymbol{v}_q$. After querying the database, we get a ranked list of images $I_1, I_2, \ldots, I_k, \ldots$ for the query $q$, where $I_k$ is the $k$-th top ranked image. Similarly, we denote the global image vector of $I_k$ as $\boldsymbol{v}_k$.

We consider the query as the heat source, and re-rank top-ranked $k$ images by computing their temperatures with the linear anisotropic diffusion equation Eq.(\ref{heat_equation1}).
For simplicity, as performed in the previous section, we also use the following more specific equation:
\begin{equation}\label{eq:re-ranking}
\boldsymbol{\mu}_q = (\boldsymbol{I}_{k}-\boldsymbol{\Lambda}^{-1}_{q}\boldsymbol{Q}_2)^{-1}(\boldsymbol{\Lambda}^{-1}_{q}\boldsymbol{Q}_1),
\end{equation}
to compute temperature gains of each image, where $\boldsymbol{Q}_2 \in \mathbb{R}^{k\times k}$ contains the cosine similarity scores among image vectors $\boldsymbol{v}_1, \ldots, \boldsymbol{v}_k$,
and $\boldsymbol{Q}_1 \in \mathbb{R}^{k\times 1}$ represents the similarity vector with the $m$-th entry storing the similarity between $\boldsymbol{v}_q$ and $\boldsymbol{v}_m$ ($m=1, 2, \ldots, k$).\footnote{In practice, before computing the similarity between any two different vectors, we first center the set of image vectors. That is, $\boldsymbol{v}_q := \boldsymbol{v}_q- \frac{1}{k+1}(\boldsymbol{v}_q+\sum_{i=1}^{k}\boldsymbol{v}_i)$, $\boldsymbol{v}_m := \boldsymbol{v}_m- \frac{1}{k+1}(\boldsymbol{v}_q+\sum_{i=1}^{k}\boldsymbol{v}_i)$. }
Additionally,  $\boldsymbol{\Lambda}_{q}$ is the diagonal matrix that is similar to $\boldsymbol{\Lambda}_{l}$ appeared in Eq.(\ref{heat_equation3}), and $\boldsymbol{\mu}_q \in \mathbb{R}^{k\times 1}$ with the $m$-th entry $\mu_q(m)$  denoting the temperature gain of $I_m$.
Apparently, we re-rank $I_1, I_2, \ldots, I_k$ based on $\boldsymbol{\mu}_q$, and images with larger temperature gains are ranked higher.

The additional computational burden of both memory usage and running time introduced by HeR is negligible. In fact, HeR is computed on image-level descriptors, and it only refines a shortlist of the top $k$ best results.
This means we only need to store $k+1$ image vectors (query as well as $k$ results), and matrices of size $k \times k$ contained in Eq.(\ref{eq:re-ranking}). In practice, both image vector dimensions and $k$ are relatively small (a few hundreds), showing the low memory usage of HeR. Furthermore, computing $\boldsymbol{\mu}_q$ is very fast with $k=800$ (adopted in our experiments), and we observe that the actual computing time on our platform is less than 30ms.

\section{Experiments and Results}\label{sec:experiments}
This section describes the implementation of our method, and reports results on public image retrieval benchmarks\footnote{Our code is available at https://github.com/MaJinWakeUp/HeWR.}.
Throughout the section, we normalize the final image vector to have unit Euclidean norm.

\subsection{Datasets and Evaluation Protocol}
We evaluate our method on three public datasets: \textbf{Oxford Buildings} (Oxford5k)~\cite{philbin2007object}, \textbf{Paris} (Paris6k)~\cite{Philbin08lost}, and \textbf{INRIA Holidays} (Holidays)~\cite{jegou2010improving}.

\vspace{0.01in}
\textbf{Oxford5k} contains a set of 5,062 photographs comprising 11 different Oxford landmarks.
There are 55 query images with each 5 queries corresponding to a landmark.
The ground truth similar images with respect to each query is provided by the dataset.
Following the standard protocol, we crop the query images based on the provided bounding boxes before retrieval.
The performance is measured using mean average precision (mAP) over the 55 queries,
where \emph{junk} images are removed from the ranking.

\vspace{0.01in}
\textbf{Paris6k} consists of 6,392 high resolution images of the city Paris.
Similar to Oxford5k, it was collected images from Flickr by querying the associated
text tags for famous Paris landmarks.
Additionally, this dataset also provides 55 query images and their corresponding ground truth relevant images.
We also use cropped query images to perform retrieval, and measure the overall retrieval performance using mAP.

\vspace{0.01in}
\textbf{Holidays} includes 1,491 images in total,
and selects 500 images as queries associated with the 500 partitioning groups of the image set. To be directly comparable with recent works~\cite{babenko2015aggregating,kalantidis2016cross,hoang2017selective},
we manually fix images in the wrong orientation by rotating them by $\pm 90$ degrees.
The retrieval quality is also measured using mAP over 500 queries, with the query removed from the ranked list.

\vspace{0.01in}
\textbf{Flickr100k}~\cite{philbin2007object} was crawled from Flickr's 145 most popular tags and consists of 11,071 images.
We combine these 100k distractor images with Oxford5k and Paris6k, and produce Oxford105k and Paris106k datasets respectively.
In this way, we evaluate the behavior of our method at a larger scale.

\begin{table*}[t]
\centering
\caption{Computational cost for the considered combinations on Oxford105k. We measure image representation time with images of size $1,024 \times 768$ (the number of features is typically $3,072$). For image representation time, we do not include time for feature extraction.}\label{times}
\begin{tabular}{|l|c|c c c c c|}
  \hline
  \multirow{2}{*}{Method}&\multirow{2}{*}{Representation time}   & \multicolumn{5}{c|}{Query time with varied image vector dimensions}\\
  \cline{3-7}
  &  &32  &64  &128 &256  &512  \\
  \hline
  SumA&13ms  &\multirow{2}{*}{8ms}  &\multirow{2}{*}{9ms}  &\multirow{2}{*}{11ms} &\multirow{2}{*}{13ms}  &\multirow{2}{*}{15ms}  \\
  HeW &173ms   & & & & & \\
  \hline
  SumA+QE+HeR&13ms &\multirow{2}{*}{45ms} &\multirow{2}{*}{47ms} &\multirow{2}{*}{48ms} &\multirow{2}{*}{50ms}  &\multirow{2}{*}{52ms}  \\
  HeW+QE+HeR&173ms & & & & & \\
  \hline
\end{tabular}
\end{table*}
\subsection{Implementation Notes}

\textbf{Deep convolutional features.} In order to extensively evaluate our method,
we use two pre-trained and a fine-tuned deep neural networks to extract multiple deep convolutional features for each image.
The adopted pre-trained networks are VGG16~\cite{simonyan2014very} and ResNet50~\cite{he2016deep}, which are widely used in the literature.
The fine-tuned network is siaMAC$\dagger$~\cite{radenovic2016cnn}, a popular fine-tuned model of VGG16.

Following the practice of previous works~\cite{kalantidis2016cross,radenovic2016cnn},
we choose the last convolutional layer of each network to separately extract patch-level image features.
We use public available trained models. Specifically, we use the MatConvNet toolbox~\cite{vedaldi2015matconvnet} for VGG16 and ResNet50,
and use the model provided in~\cite{arandjelovic2016netvlad}  for siaMAC$\dagger$. In addition, in order to accelerate feature extraction,
we resize the longest side of all images to 1,024 pixels while preserving aspect ratios before feeding them  into each deep network.

\vspace{0.01in}
\textbf{PCA whitening} is widely used in many image retrieval systems~\cite{jegou2012negative,kalantidis2016cross,hoang2017selective,babenko2015aggregating} as it can effectively improve the discriminative ability of image vectors.
In order to avoid over-fitting,
the PCA matrix is usually learned with the set of aggregated image vectors produced from a held-out dataset.
To be directly comparable with related works, we learn PCA parameters on Paris6k for Oxford5k and Oxford105k,
and on Oxford5k for Paris6k and Paris106k.
As for Holidays, we randomly select a subset of 5,000 images from Flickr100k to learn parameters.

\vspace{0.01in}
\textbf{Query Expansion (QE)}~\cite{chum2007total} is an effective post-processing technique to increase retrieval performance.
Given the ranked list of database images over a query image, we simply calculate the average vector of the 10 top-ranked image vectors and the query vector, and we then use the $L_2$ normalized average vector to re-query again.
After QE, we then apply our re-ranking algorithm HeR to further improve retrieval performance.
We will show the combination of QE and HeR is beneficial in practice.

\subsection{Impact of the Parameters} 
We investigate the impact of the parameter $k$ as well as the impact of the final image vector dimensionality
to different retrieval frameworks with deep convolutional features extracted by the network siaMAC$\dagger$ on the datasets of Oxford5k and Oxford105k.
Specifically, we evaluate their impact to
the following four combinations: SumA, HeW, SumA+QE+HeR and HeW+QE+HeR.
SumA means we obtain image representations by simply setting $w_l \equiv 1$ in Eq.(\ref{representation_equation}), and perform image search by linearly scanning database vectors.
It justifies the contribution of our deep feature aggregation method HeW.
SumA+QE+HeR and HeW+QE+HeR indicate that we perform re-ranking QE+HeR with image vectors produced by SumA and HeW respectively.
We use them to determine the parameter $k$ that is introduced in HeR.
Meanwhile, by setting $k=0$, they can also be used for evaluating the effect of our image re-ranking strategy HeR.

\begin{figure}[t]
\centering
\includegraphics[width=1.72in]{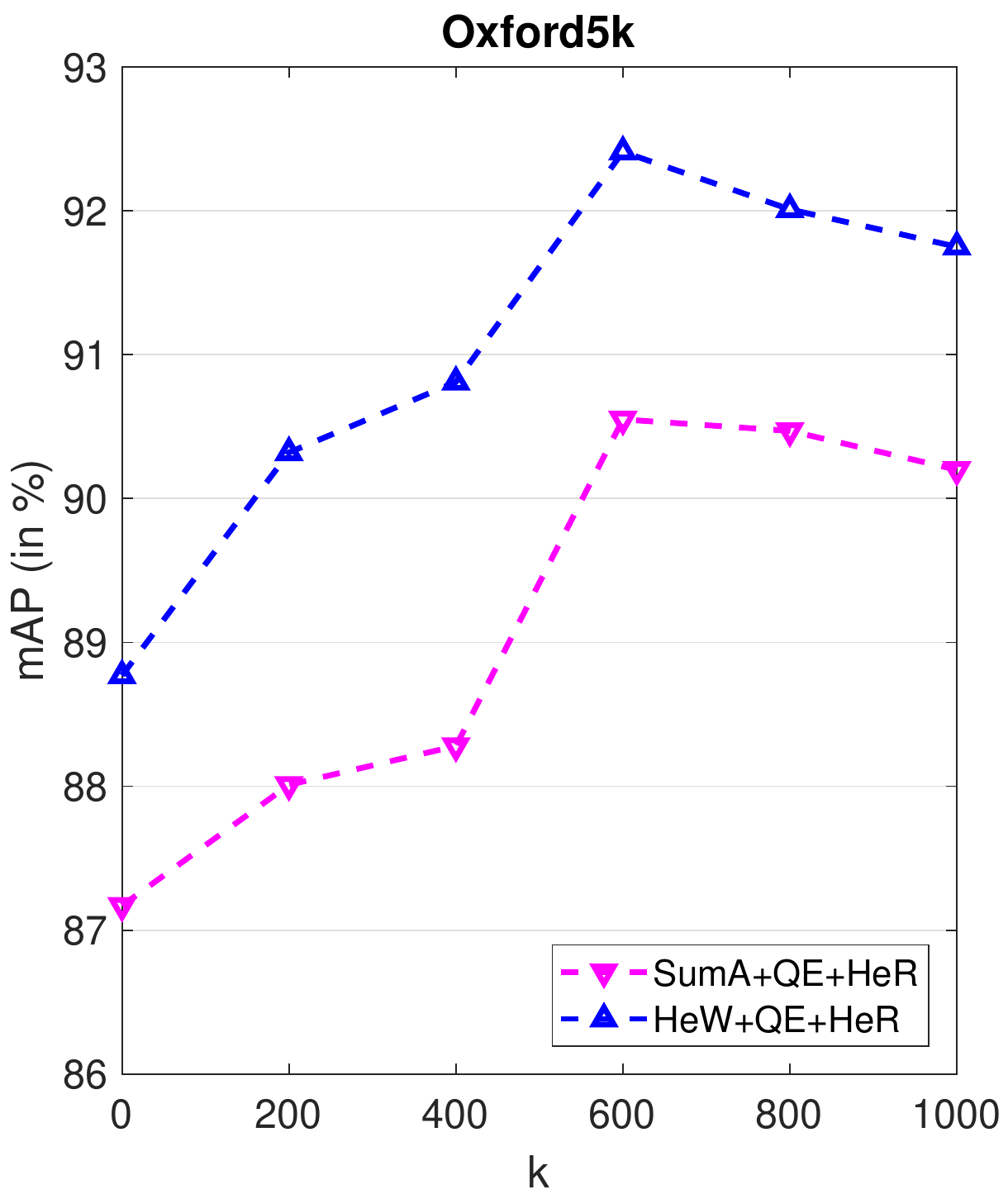}
\includegraphics[width=1.72in]{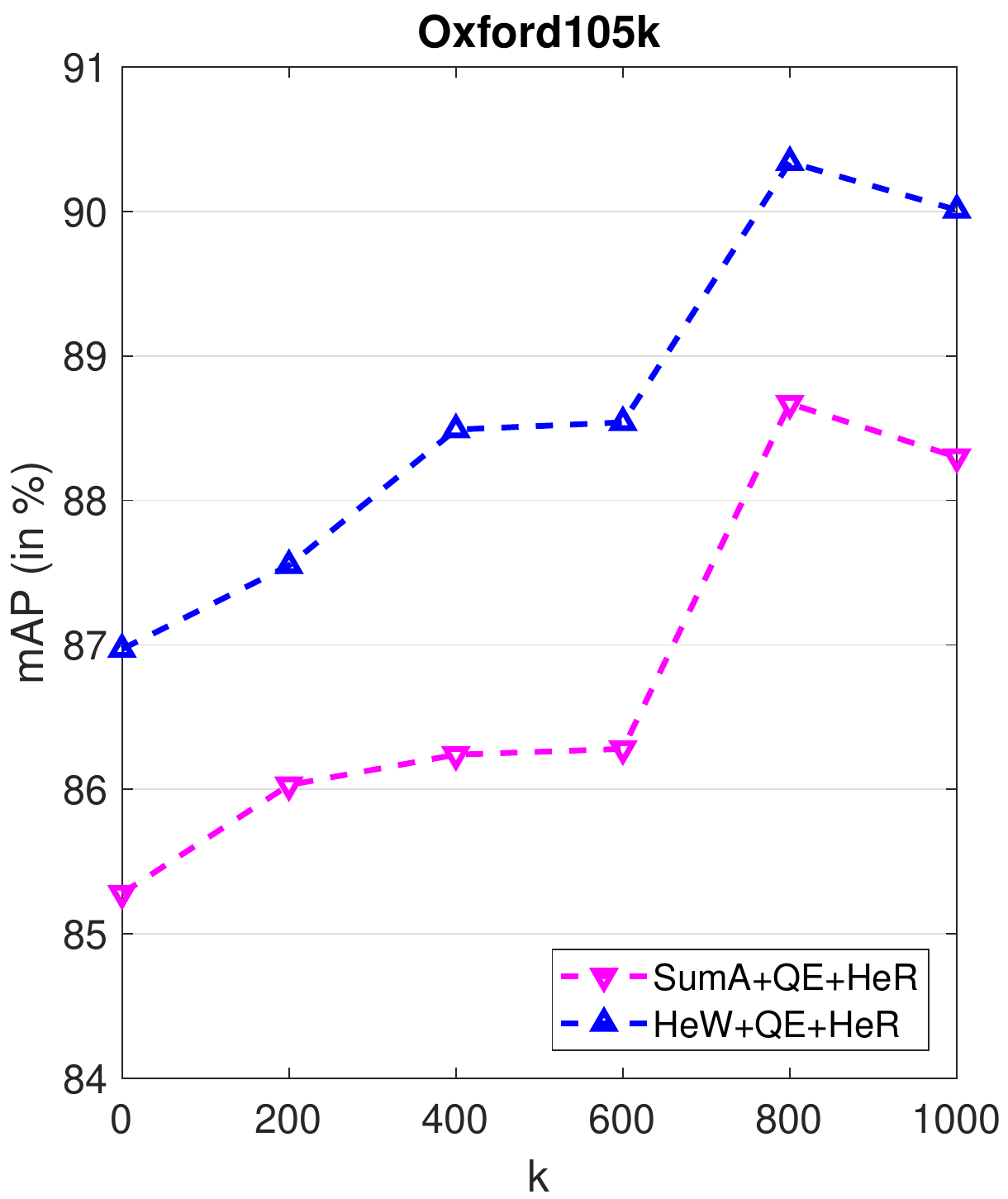}
\caption{Impact of the number of re-ranking images to retrieval quality. The image vector dimensions $D=512$, and it is worth noting that  HeW+QE+HeR (SumA+QE+HeR) degenerates to  HeW+QE (SumA+QE) when $k=0$.}
\label{fig:k}
\end{figure}

\begin{table*}[t]
\begin{center}
\caption{Impact of the networks to the considered methods. We do not perform re-ranking on Holidays as as it is not a standard practice.}\label{comparison_with_baseline}
\begin{tabular}{|l|c|c| c c c c c|}
  \hline
  Network                    &Method             &~Dim.~  &Oxford5k  &Oxford105k  &Paris6k  &Paris106k &Holidays \\
  \hline
  \hline
  \multirow{10}{*}{ResNet50} &SumA               &2,048    &71.7  &65.9    &83.0     &76.8          &89.4\\
                             &HeW                &2,048    &72.1  &66.2    &84.5     &78.6          &\textbf{90.1}  \\
                             &SumA+QE            &2,048    &76.2  &71.9    &87.8     &81.2          &-- \\
                             &HeW+QE             &2,048    &77.2  &73.4    &89.7     &83.5          &--  \\
                             &HeW+QE+HeR         &2,048    &80.1  &75.4    &91.0     &85.1          &--  \\
                             \cline{2-8}
                             &SumA               &512     &59.6  &52.8    &81.1     &72.8          &88.1 \\
                             &HeW                &512     &61.1  &54.4    &82.9     &75.0          &88.3  \\
                             &SumA+QE            &512     &61.5  &55.3    &84.2     &77.8          &-- \\
                             &HeW+QE             &512     &63.2  &57.5    &86.5     &80.2          &--  \\
                             &HeW+QE+HeR         &512     &67.7  &62.0    &90.1     &82.3          &-- \\
  \hline
  \multirow{5}{*}{VGG16}     &SumA               &512     &71.0  &66.0        &80.6     &73.6      &87.8  \\
                             &HeW                &512     &72.8  &68.0        &81.5     &74.4      &88.4  \\
                             &SumA+QE            &512     &76.5  &74.3        &85.5     &81.5      &--  \\
                             &HeW+QE             &512     &77.8  &75.3        &87.0     &82.7      &--  \\
                             &HeW+QE+HeR         &512     &82.0  &78.7        &91.2     &86.1      &--  \\
  \hline
  \multirow{5}{*}{SiaMAC$\dagger$} &SumA         &512    &80.8  &76.5        &86.3     &80.6      &86.3 \\
                             &HeW                &512    &82.6  &78.8        &87.0     &81.3      &87.1  \\
                             &SumA+QE            &512    &87.3  &85.2        &89.4     &85.2      &--  \\
                             &HeW+QE             &512    &88.8  &87.0        &90.7     &86.6      &--  \\
                             &HeW+QE+HeR         &512    &\textbf{92.0}  &\textbf{90.3}        &\textbf{94.3}     &\textbf{90.2}   &--\\
  \hline
\end{tabular}
\end{center}
\end{table*}

\vspace{0.01in}
\textbf{Impact of the number of re-ranking images.} We first use the retrieval frameworks SumA+QE+HeR and HeW+QE+HeR to evaluate the impact of $k$ to retrieval quality with full image vector dimensions $D=512$.
The mAP performance for the considered two frameworks on Oxford5k and Oxford105k under different $k$ values is shown in Fig.~\ref{fig:k}, where $k=0$ means HeR is not applied.
As we see,  HeR consistently improves the retrieval quality, and the margin is around 3\% for the  two largest $k$ values.
The best results for HeW+QE+HeR and SumA+QE+HeR on Oxford5k and Oxford105k are achieved at $k=600$ and $k=800$,
respectively. Since $k=800$ gives overall better results than $k=600$, therefore it is used in the following experiments.

\begin{figure}[t]
\centering
\includegraphics[width=1.72in]{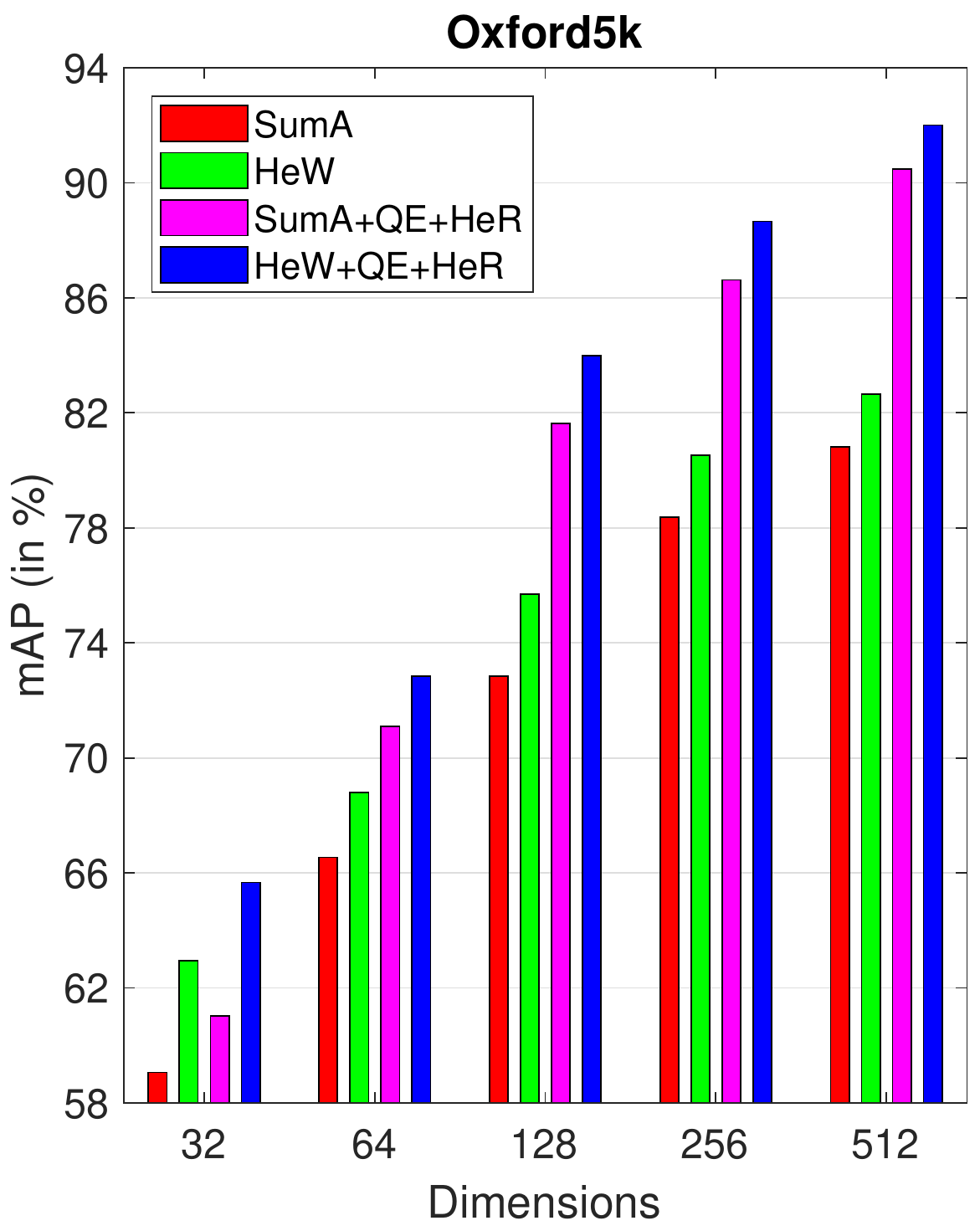}
\includegraphics[width=1.72in]{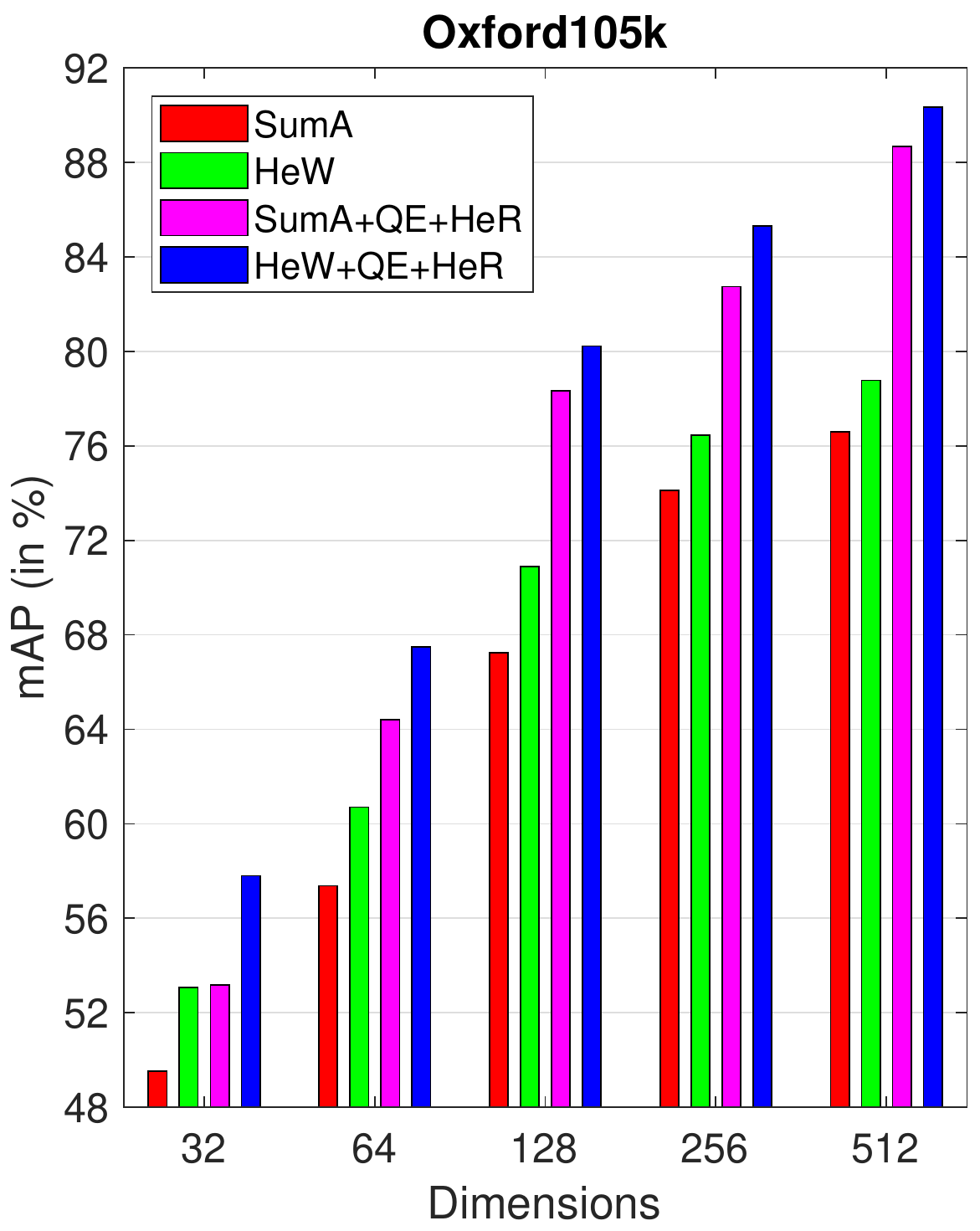}
\caption{Impact of the the final image vector dimensionality to the considered combinations. Short representations are achieved by keeping only the first components of 512 dimensional image vectors after PCA whitening.}
\label{fig:dimensions}
\end{figure}

\vspace{0.01in}
\textbf{Impact of the final image vector dimensionality.} With $k=800$, we illustrate mAP curves when varying the dimensionality of the final image vectors from 32 to 512 in Fig.~\ref{fig:dimensions}.
Dimensionality reduction is achieved by keeping only the first $D$ components of 512 dimensional image vectors after PCA whitening.
As illustrated, the gain of HeW over SumA is nearly 2\% in mAP on both Oxford5k and Oxford105k at $D= 512$.
Furthermore, the performance gain is increasing with the reduction of the number of dimensions, and the gain is around 4\% at $D= 32$.
This means our image representation method HeW affects less by dimensionality reduction than the baseline SumA.

Both HeW and SumA are significantly benefited with image re-ranking, especially with image vectors of higher dimensions.
As shown, when $D=512$, the increased mAP values of HeW+QE+HeR (SumA+QE+HeR) over HeW (SumA) on Oxford5k and Oxford105k are 9.4\% (9.7\%) and 11.5\% (12.1\%), respectively.
After incorporating re-ranking, the performance advantage of HeW over SumA is enlarged at low dimensions.
For instance, the increased mAP values of HeW+QE+HeR over SumA+QE+HeR on Oxford5k and Oxford105k at $D= 32$  are 4.7\% and 4.6\%, respectively.

\vspace{0.01in}
\textbf{Computational cost.}
We now turn to present running time for the considered combinations on Oxford105k. Table~\ref{times} reports timings (excluding time for feature extraction) measured to compute image representations and to perform image querying. We implement both SumA and HeW in Matlab, and benchmarks are obtained on an Intel Xeon E5-2630/2.20GHz with 20 cores.
As the table shows, although the baseline method SumA is  faster than HeW by about an order of magnitude, HeW is still fast in practice.
For an image of high resolution $1,024 \times 768$, the number of deep features is $3,072$,  and the time to derive image vector with HeW is typically 173ms. It is worth noting that, the number of features extracted by siaMAC$\dagger$ is four times of that extracted by VGG16 and ResNet50.
This means aggregating features produced VGG16 and ResNet50 is much faster.
In practice, we observe that, we can accomplish feature aggregation in less than 10ms with features extracted by both VGG16 and ResNet50.

It is simple to know that HeW does not affect search efficiency, and it has the same searching time as SumA. Thus, the online query time for HeW is about 188ms at $D=512$.
The increased time caused by incorporating image re-ranking into the retrieval system is very limited. As shown, the total online processing time for HeW+QE+HeR with $D=512$ is only about 225ms. This means searching with our method is efficient in practice.

\begin{table*}[t]
\begin{center}
\caption{Performance (in mAP) comparison with methods using SIFT and off-the-shelf available networks.
As many previous works, we do not perform re-ranking on Holidays as each query only has a few ground truth similar images.
}\label{comparison_with_unsupervised}
\begin{tabular}{|c|l|r|c c c c c|}
  \hline
  Feature &Method                                 &~Dim.~  &Oxford5k  &Oxford105k  &Paris6k  &Paris106k &Holidays \\
  \hline
  \hline
  \multirow{2}{*}{{SIFT}}
   &F-FAemb~\cite{do2017embedding}         &7,245 &66.1      &64.3        &--       &--        &75.5     \\
   &Temb~\cite{murray2017interferences}    &8,064 &70.0      &64.4        &--       &--        &71.6     \\
  \hline
  Fully&MOP-CNN~\cite{gong2014multi}           &2,048   &--        &--          &--       &--        &80.8      \\
  connected &Neural codes~\cite{babenko2014neural}  &4,096   &54.5      &51.2        &--       &--        &79.3        \\
  layer&CNNaug-ss~\cite{razavian2014cnn}       &4,096 &68.0      &--          &79.5     &--        &84.3   \\
  \hline
  \multirow{13}{*}{\rotatebox{90}{Deep Conv. layer of VGG16}} 
  &R-MAC~\cite{tolias2016particular}      &512   &66.9      &61.6        &\textbf{83.0}     &75.7 &-- \\
  &CroW~\cite{kalantidis2016cross}        &512   &70.8      &65.3        &79.7     &72.2      &85.1      \\
  &SUM-mask~\cite{hoang2017selective}     &512   &64.0      &58.8        &78.6     &70.4      &86.4      \\
  &MAX-mask~\cite{hoang2017selective}     &512   &65.7      &60.5        &81.6     &72.4      &85.0      \\
  &PWA~\cite{xu2018unsupervised}          &512   &72.0      &66.2        &82.3     &\textbf{75.8}  &--  \\
  &ReSW~\cite{pang2018building}           &512   &72.6      &67.5        &82.4     &73.0      &85.3   \\
  &HeW                                    &512   &\textbf{72.8}  &\textbf{68.0}        &81.5     &74.4  &\textbf{88.4} \\
  \cline{2-8}
  &R-MAC+AML+QE~\cite{tolias2016particular}&512   &77.3 &73.2   &86.5     &79.8 &-- \\
  &CroW+QE~\cite{kalantidis2016cross}     &512   &74.9      &70.6        &84.8     &79.4       &--\\
  &PWA+QE~\cite{xu2018unsupervised}       &512   &74.8      &72.5        &86.0     &80.7      &--       \\
  &ReSW+QE~\cite{pang2018building}        &512   &76.3      &73.5        &86.3     &80.2      &--   \\
  &HeW+QE                                 &512   &77.8      &75.3        &87.0     &82.7      &-- \\
  &HeW+QE+HeR                             &512   &\textbf{82.0}   &\textbf{78.7}    &\textbf{91.2}           &\textbf{86.1}   &--\\
  \hline
\end{tabular}
\end{center}
\end{table*}

\subsection{Impact of the Networks}
Table~\ref{comparison_with_baseline}  illustrates the impact of the evaluated networks to the baseline and our methods.
Although ResNet50 has demonstrated much superior performance than VGG16 on the ILSVRC classification task~\cite{he2016deep,vedaldi2015matconvnet}, it does not actually produce better performance than the latter on the image retrieval benchmarks. As shown, while relying on much higher image representation dimensions $D=2,048$, ResNet50-based results are still inferior to VGG16-based results in many cases.
Even worse, when reducing the dimensionality to 512 components using PCA, ResNet50-based results fall behind the corresponding VGG16-based results by large margins except on the dataset of Holidays.
As is expected, SiaMAC$\dagger$-based results outperform VGG16-based results on the datasets of Oxford and Paris as the network SiaMAC$\dagger$ is fine-tuned with a large number of landmark building photos.
However, it is worth noting that, fine-tuning may result in over-fitting.
As shown, after fine-tuning, the performance for both SumA and HeW is slightly decreased on the Holidays.

Both the proposed feature aggregation method HeW and image re-ranking method HeR give boost in performance.
As shown in Table~\ref{comparison_with_baseline}, although HeW outperforms SumA by a little margin in some cases, it outperforms the latter by over 1\% in mAP in most cases after incorporating QE. Furthermore, we obtain additional 3\% mAP gains with our re-ranking method HeR. Accordingly, the proposed complete method HeW+QE+HeR typically outperforms the baseline method SumA+QE by 4\% with the adopted networks.

\begin{table*}[t]
\begin{center}
\caption{Performance (in mAP) comparison with the state-of-the-art methods using unsupervised fine-tuned networks. }\label{comparison_with_fine_tuning}
\begin{tabular}{|l|c| c c c c c|}
  \hline
  Method                                 &~Dim.~  &Oxford5k  &Oxford105k  &Paris6k  &Paris106k &Holidays \\
  \hline
  \hline
  NetVLAD~\cite{arandjelovic2016netvlad} &512 &67.6  &--          &74.9     &--        &86.1\\
  siaMAC$\dagger$+MAX-mask~\cite{hoang2017selective}&512   &77.7      &72.7        &83.2     &76.5      &86.3\\
  Fisher Vector~\cite{ong2017siamese}               &512   &81.5  &76.6      &82.4        &--       &--   \\
  siaMAC$\dagger$+MAC~\cite{radenovic2016cnn}     &512   &79.0      &73.9        &82.4     &74.6      &79.5\\
  siaMAC$\dagger$+R-MAC~\cite{radenovic2016cnn}     &512   &77.0      &69.2        &83.8     &76.4      &82.5\\
  siaMAC$\dagger$+ReSW~\cite{pang2018building}     &512   &\textbf{83.4}      &\textbf{79.3}        &86.5     &80.1      &85.5\\
  siaMAC$\dagger$+HeW                              &512   &82.6 &78.8 &\textbf{87.0} &\textbf{81.3} &\textbf{87.1}    \\
  \hline
  siaMAC$\dagger$+MAC~\cite{radenovic2016cnn}       &256   &77.4      &70.7       &80.8     &72.2      &77.3\\
  siaMAC$\dagger$+R-MAC~\cite{radenovic2016cnn}     &256   &74.9      &67.5       &82.3     &74.1      &81.4\\
  siaMAC$\dagger$+HeW                              &256    &80.5      &76.5       &85.7     &79.9      &85.9\\
  \hline
  siaMAC$\dagger$+MAC~\cite{radenovic2016cnn}       &128   &75.8      &68.6        &77.6     &68.0      &73.2\\
  siaMAC$\dagger$+R-MAC~\cite{radenovic2016cnn}     &128   &72.5      &64.3        &78.5     &69.3      &79.3\\
  siaMAC$\dagger$+HeW                              &128    &75.7      &70.9        &81.7     &75.0      &85.0\\
  \hline
  \hline
  siaMAC$\dagger$+MAC+R+QE~\cite{radenovic2016cnn}       &512   &85.0  &81.8   &86.5    &78.8    &--\\
  siaMAC$\dagger$+R-MAC+R+QE~\cite{radenovic2016cnn}     &512   &82.9  &77.9   &85.6    &78.3    &--\\
  siaMAC$\dagger$+HeW+QE                                &512   &\textbf{88.8}  &\textbf{87.0}   &\textbf{90.7}  &\textbf{86.6 }  &-- \\
  siaMAC$\dagger$+HeW+QE                                &256   &85.8  &83.9   &89.6  &85.5  &-- \\
  siaMAC$\dagger$+HeW+QE                                &128   &80.7  &78.2   &86.4  &81.6  &-- \\
  \hline
  \hline
  siaMAC$\dagger$+HeW+QE+HeR                   &512    &\textbf{92.0}  &\textbf{90.3}     &\textbf{94.3}     &\textbf{90.2}     &--\\
  siaMAC$\dagger$+HeW+QE+HeR                   &256    &88.7   &85.3     &93.0     &88.2     &--\\
  siaMAC$\dagger$+HeW+QE+HeR                   &128    &84.0   &80.2     &88.4     &84.3    &--\\
  \hline
\end{tabular}
\end{center}
\end{table*}

\begin{figure*}[t]
\centering
\includegraphics[width=5.5in]{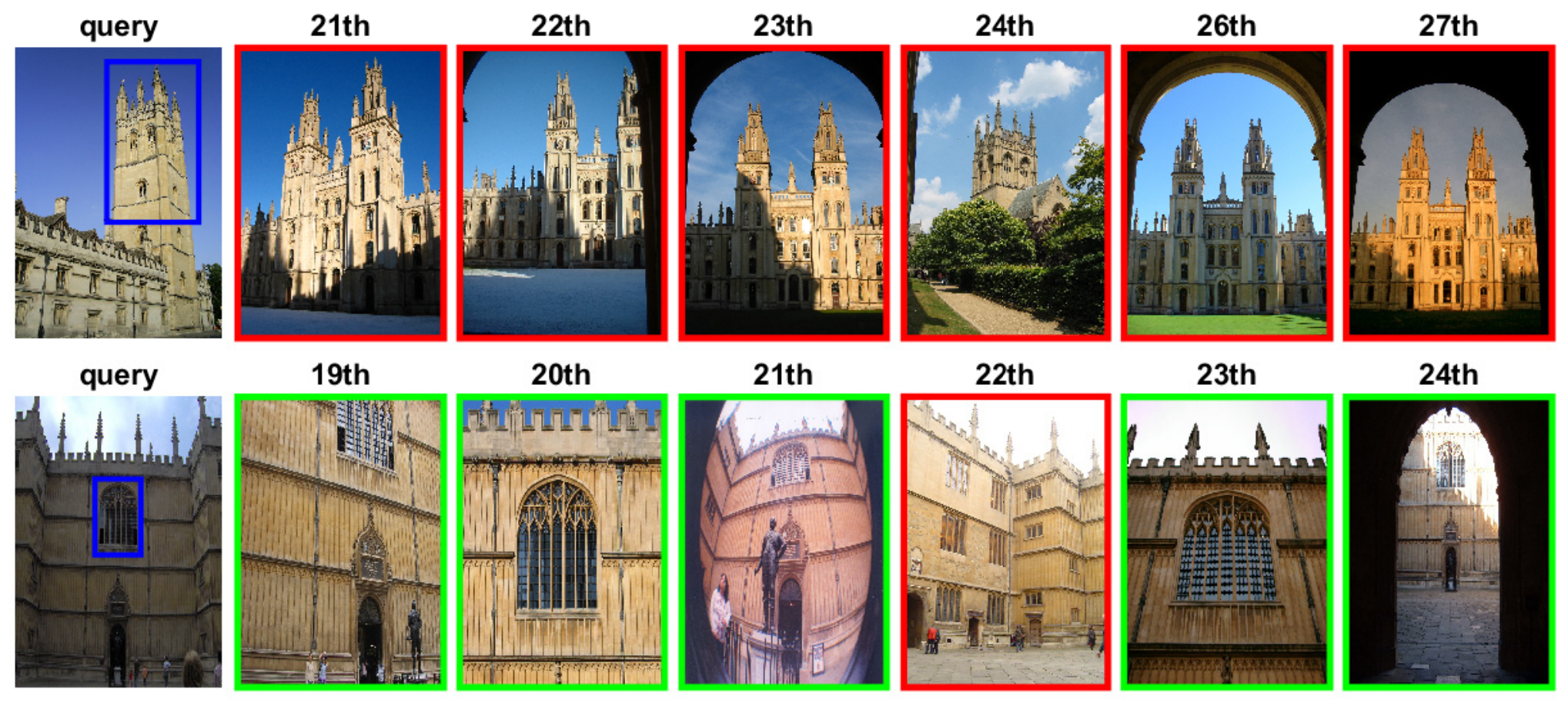}
\caption{A failure example (top) and a successful example (bottom) from the test set Oxford105k for our complete approach HeW+QE+HeR.
There are 27 false positive images ranked between 21 and 54 (the number of ground truth similar images is 54) for the failure case.
For this example, its average precision is only 58.4\% and its top six false positive results are displayed for illustration.
For the successful example, there are 24 ground truth similar images, and there is only one false positive image ranked higher than 24.
As illustrated, we show its retrieval results ranked from 19th to 24th. Note, the blue/red/green border represent query region/false positive result/true positive result, respectively.}
\label{fig:ox105k_qer}
\end{figure*}

\subsection{Comparison with the State-of-the-art}
We below show the comparison results of the proposed approach with related
unsupervised methods that use off-the-shelf  and fine-tuned networks separately.

\vspace{0.01in}
\textbf{Comparison with methods using SIFT and pre-trained networks.}
In Table~\ref{comparison_with_unsupervised}, we present comparisons of our approach using VGG16 with methods using SIFT and off-the-shelf available networks, which utilize global representations of images. The comparison results are summarized as follows:
\begin{itemize}
  \item Our approach HeW significantly outperforms two state-of-the-art methods~\cite{do2017embedding,murray2017interferences} using weaker SIFT features, although their dimensions are more than 10 times higher than ours.
  Furthermore, it also shows clear advantages over~\cite{gong2014multi,babenko2014neural,razavian2014cnn} that utilize fully connected layers to derive image representations.
  \item Compared with~\cite{tolias2016particular,kalantidis2016cross,hoang2017selective,xu2018unsupervised,pang2018building} which are also using the VGG16 model, our method HeW achieves the best results on the datasets of Oxford5k, Oxford105k and Holidays. Additionally, it is worth noting that, the gain on Holidays is over 3\% at the same dimensionality.
  \item When combined with query expansion, our approach outperforms the compared methods~\cite{tolias2016particular,kalantidis2016cross,xu2018unsupervised,pang2018building} on all evaluated datasets. It should be noted that~\cite{tolias2016particular} includes a nontrivial and computationally intensive spatial verification process.
  \item Applying HeR after QE results in significant performance gains. As shown, our re-ranking strategy HeR gives a boost of around 3.5\% in mAP on all evaluated datasets. Consequently, we outperform the compared methods on two large-scale datasets Oxford105k and Paris106k by more than 5\% in mAP.
\end{itemize}

\begin{table*}[t]
\begin{center}
\caption{Comparison with the best results reported in the literature. }\label{comparison_with_best}
\begin{tabular}{|l|r| c c c c c|}
  \hline
  Method                                          &~~Dim.  &Oxford5k  &Oxford105k  &Paris6k  &Paris106k &Holidays \\
  \hline
  \hline
  Mikulik \textit{et al.}~\cite{mikulik2013learning}       &16M        &84.9    &82.4     &79.5     &77.3   &75.8\\
  Tolias \textit{et al.}~\cite{tolias2013aggregate}        &8M        &87.9    &--       &85.4     &--     &85.0\\
  Tolias and J{\'e}gou~\cite{tolias2015visual}    &8M        &89.4    &84.0     &82.8     &--     &--  \\
  Arandjelovic \textit{et al.}~\cite{arandjelovic2016netvlad}&4,096 &71.6    &--       &79.7     &--     &87.5\\
  Hoang \textit{et al.}~\cite{hoang2017selective}          &4,096   &83.8    &80.6     &88.3     &83.1   &\textbf{92.2}\\
  Iscen \textit{et al.}~\cite{iscen2017efficient}          &5$\times$512   &91.5    &84.7     &95.6     &\textbf{93.0}    &--\\
  Iscen \textit{et al.}~\cite{iscen2017fast}               &5$\times$512   &91.6    &86.5     &95.6    &92.4 &--\\
  Noh \textit{et al.}~\cite{noh2017largescale}             &--             &90.0    &88.5     &\textbf{95.7}    &92.8  &--\\
  Gordo \textit{et al.}~\cite{gordo2016deep}                &512     &89.1    &87.3     &91.2     &86.8    &89.1\\
  \hline
  This paper                                        &512    &\textbf{92.0}  &\textbf{90.3}     &94.3     &90.2     &88.4\\
  \hline
\end{tabular}
\end{center}
\end{table*}

\vspace{0.01in}
\textbf{Comparison with methods using  fine-tuned networks.}
We perform comparisons with recent unsupervised fine-tuned methods~\cite{arandjelovic2016netvlad,radenovic2016cnn,hoang2017selective,ong2017siamese,pang2018building} in Table~\ref{comparison_with_fine_tuning}.
The table again demonstrates the superior performance of our approach over related baselines at the same dimensionality:
\begin{itemize}
  \item Although HeW slightly falls behind our very recent work~\cite{pang2018building} on Oxford5k and Oxford105k, we establish new state-of-the-art results on the other three evaluated datasets at dimensionality of 512, and the improved mAP values over~\cite{arandjelovic2016netvlad,radenovic2016cnn,hoang2017selective,ong2017siamese} are
      not negligible. For example, the gain on Paris106k is at least 4.7\%.
  \item With the same siaMAC$\dagger$ features, our approach improves two related baselines R-MAC and MAC presented in~\cite{radenovic2016cnn} without and with dimensionality reduction, and the improvement is more significant on two large datasets Oxford105k and Paris106k.
  \item Our method, HeW, also produces better results than~\cite{radenovic2016cnn} after query expansion,
  although~\cite{radenovic2016cnn} uses more sophisticated post-processing than us.
  Additionally, HeW outperforms ~\cite{radenovic2016cnn} at $D=256$, and keeps competitive even at $D=128$.
  \item Finally, we further enlarge the mAP gains over the compared methods by applying HeR after QE. As shown, at  $D=512$,
  we outperform~\cite{radenovic2016cnn} by 7.0\%, 8.5\%, 7.8\%, 11.4\% on Oxford5k, Oxford105k, Paris6k, Paris106k, respectively.
\end{itemize}

To better understand our complete method HeW+QE+HeR, we visualize two example query images in Fig.~\ref{fig:ox105k_qer}.
The top query example can be considered as a failure case for our method as its average precision is only 58.8\%, falling far behind the mAP value of 92.0\%. As displayed, although the landmark contained in the 24-th ranked image is exactly not the same as the one contained in the query, it is visually similar to the query.
For the bottom query example, there are 24 ground truth similar images, and there is only one false positive image ranked at 22-th.
Its average precision is 97.0\%, and thus it can be seen as a successful example.
As shown, the unique false positive image contains several window patches, and therefore it is understandable that it has a large similarity score with the query region.

\vspace{0.01in}
\textbf{Comparison with costly methods.}
Table~\ref{comparison_with_best} compares our best results
with costly methods that focus on spatial verification or matching kernel.
Some of them~\cite{iscen2017efficient,iscen2017fast,noh2017largescale} do not necessarily rely on a
global representation, and some others~\cite{mikulik2013learning,tolias2013aggregate,tolias2015visual, arandjelovic2016netvlad,hoang2017selective} represent images with much higher dimensional vectors, and are thus not directly comparable.
There is no doubt that these methods use a larger memory footprint than our approach.
Additionally, their search efficiency is obviously  much lower than our method.
For instance, the method of~\cite{tolias2015visual} requires a slow spatial verification taking over 1
second per query (excluding descriptor extraction time). This means these best results are  hardly scalable as they require a lot of storage memory and searching time.
Compared with these methods, we still produce the best performance on Oxford5k and Oxford105k.
Similar to our approach, the supervised fine-tuned method~\cite{gordo2016deep} also represents images with 512 dimensional vectors.
Compared with this method, we produce slightly inferior result on Holidays, and achieve much better performance on the other datasets.

\section{Conclusions}\label{sec:conclusions}
We proposed an efficient aggregation approach for building compact but powerful image representations by utilizing the heat equation in this manuscript.
We utilized the theory of anisotropic diffusion, and assumed that graph defined by a set of deep features constitutes a heat transfer system.
By considering each deep feature as a heat source, our approach avoided over-representation of bursty features by enforcing the system temperatures  derived from all features be a constant. We provided a practical solution to derive image vectors, and demonstrated the effectiveness of our method on the task of instance-level retrieval.
Inspired by our aggregation method, we also presented a heat equation based image re-ranking method to further increase retrieval performance.
Both of feature aggregation and image re-ranking methods are unsupervised, and can be compatible with different CNNs, including pre-trained and fine-tuned networks.
Experimental results showed that we have established new state-of-the-art results on public image retrieval benchmarks using 512-dimensional vector representations.

\section*{Acknowledgments}
This work was supported by National Key Research and Development Plan 2016YFB1001004,
National Natural Science Foundation of China Grant 61603289, China Postdoctoral
Science Foundation Grant 2016M602823, and Fundamental Research Funds for the Central Universities xjj2017118.

\ifCLASSOPTIONcaptionsoff
  \newpage
\fi



\bibliographystyle{IEEEtran}
\bibliography{HeWR}
%
%
%

%

\begin{IEEEbiography}[{\includegraphics[width=1.0in, height=1.25in, clip,keepaspectratio]{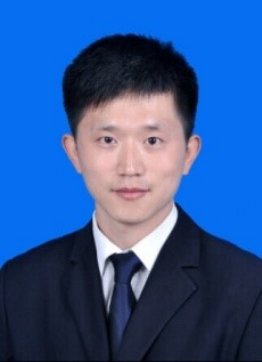}}]{Shanmin Pang} received the BS, MS, and Ph.D degrees from Shaanxi Normal University, Capital Normal University, and Xi'an Jiaotong University
in 2006, 2009 and 2015, respectively.
Prof. Pang is now an assistant professor in the School of Software Engineering at Xi'an Jiaotong University. His research interests include pattern recognition, computer vision and image processing. He won the best application paper award at the ACCV 2012 conference.
\end{IEEEbiography}

\begin{IEEEbiography}[{\includegraphics[width=1.0in, height=1.25in, clip,keepaspectratio]{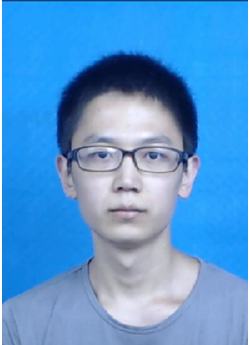}}]{Jin Ma}
received his BS degree  from Xi'an Jiaotong University, Xi'an, China, in 2016.
He is pursuing his ME degree in the School of Software Engineering at Xi'an Jiaotong University.
His research interest is computer vision, with a focus on content based image retrieval.
\end{IEEEbiography}

\begin{IEEEbiography}[{\includegraphics[width=1.0in, height=1.25in, clip,keepaspectratio]{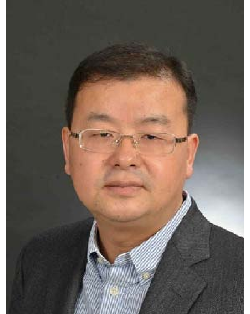}}]{Jianru Xue} (M'06)
got his BS degree from Xi'an University of Technology in 1994, and both MS and PhD degrees from Xi'an Jiaotong University in 1999
and 2003, respectively.
He is currently a full professor of the Institute of Artificial Intelligence and Robotics at Xi'an Jiaotong University, Xi'an,
China. He worked in FujiXerox, Tokyo, Japan, from 2002 to 2003, and visited University
of California, Los Angeles, from 2008 to 2009. His research field includes computer vision, visual localization and navigation, and video
coding based on analysis.
\end{IEEEbiography}

\begin{IEEEbiography}[{\includegraphics[width=1.0in, height=1.25in, clip,keepaspectratio]{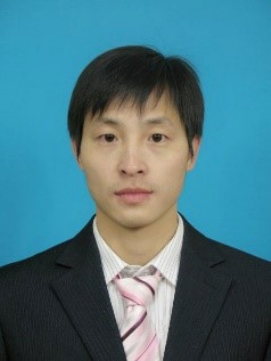}}]{Jihua Zhu}
received the BE degree in automation from Central South University, China and the Ph.D. degree in pattern recognition and intelligence systems from Xi'an Jiaotong University, China, in 2004 and 2011, respectively.
He is an associate professor in the School of Software Engineering at Xi'an Jiaotong University.
His research interests include computer vision and machine learning.
\end{IEEEbiography}

\begin{IEEEbiography}[{\includegraphics[width=1.0in, height=1.25in, clip,keepaspectratio]{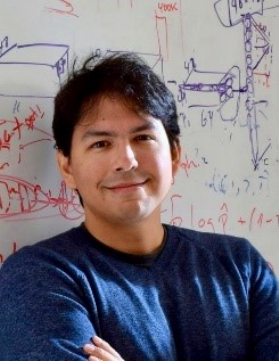}}]{Vicente Ordonez} is an assistant professor in the Department of Computer Science at the University of Virginia. His research interests lie at the intersection of computer vision, natural language processing and machine learning, with a focus on exploiting the natural connections that occur in vision and language using large amounts of data. He was a recipient of the Best Long Paper Award at EMNLP 2017, the IEEE David Marr Prize in Computer Vision in 2013. He obtained a PhD in Computer Science at the University of North Carolina at Chapel Hill, and an MS in Computer Science at Stony Brook University in New York State.
\end{IEEEbiography}




\end{document}